\documentclass[amssymb,
amsmath,
amsfonts,
reprint,
superscriptaddress,
longbibliography,
aip, jcp,
footinbib,
12pt,
onecolumn]{revtex4-2}
\usepackage[utf8]{inputenc}
\usepackage{graphicx}
\usepackage{dcolumn}
\usepackage[T1]{fontenc}
\usepackage{tabularx}
\usepackage{enumitem}
\usepackage{filecontents}
\usepackage{soul}
\usepackage{mathrsfs}
\usepackage{mathtools}
\usepackage{chemformula}

\usepackage{multirow}

\usepackage[mathlines]{lineno}
\usepackage{graphicx}
\usepackage[margin=1in]{geometry}

\usepackage[hidelinks, hypertexnames=false]{hyperref}
\definecolor{color1}{rgb}{0,0.25,0.70}
\hypersetup{colorlinks=true,
    linkcolor={color1},
    citecolor={color1},
    urlcolor={color1}
}

\newcommand{\closedm}[1]{\left[ #1 \right]}
\newcommand{\closeds}[1]{\left( #1 \right)}
\newcommand{\closede}[1]{\left\langle #1 \right\rangle}

\newcommand{\px}[1]{\hspace{#1 pt}}

\begin{document}

\title{Chiral vibrational modes and vibrational circular dichroism}
\author{Xuecheng Tao$^\dagger$}
\thanks{$^\dagger$ Equal contribution}
\email{xuechengtao@gmail.com}
\affiliation{Department of Chemistry, University of Pennsylvania, Philadelphia, Pennsylvania 19104, United States}
\author{Cl\`audia Climent$^\dagger$}
\thanks{$^\dagger$ Equal contribution}
\email{claudiacliment@ub.edu}
\affiliation{Departament de Ci\`encia de Materials i Qu\'imica F\'isica and 
Institut de Qu\'imica Te\`orica i Computacional (IQTCUB), Universitat de Barcelona, 
Martí i Franqu\`es 1, 08028 Barcelona, Spain}
\author{Ethan Abraham}
\affiliation{Department of Chemistry, Massachusetts Institute of Technology, Cambridge, Massachusetts 02139, USA}
\author{Jichen Feng}
\affiliation{Department of Chemistry, University of Pennsylvania, Philadelphia, Pennsylvania 19104, United States}
\author{Abraham Nitzan}
\email{anitzan@sas.upenn.edu}
\affiliation{Department of Chemistry, University of Pennsylvania, Philadelphia, Pennsylvania 19104, United States}
\affiliation{Department of Physical Chemistry, Tel Aviv University, Tel Aviv 6997801, Israel}

\date{\today}

\begin{abstract}
\clearpage
The recent interest in chiral phonons in a variety of physical phenomena and their hypothesized role in the chiral-induced spin selectivity effect [Phys. Rev. Research, \textbf{5}, L022039 (2023)] call for further investigation into the chirality of molecular vibrations. Although molecular chirality has conventionally been quantified using structural properties, recent work has highlighted the role of dynamical response properties as chirality metrics. In this work, we examine an inter-atom helicity pseudoscalar as a complementary measure of vibrational chirality, associated with the vibrational circular dichroism (VCD) intensity in the fixed partial charge (FPC) approximation. This pseudoscalar is translationally and rotationally invariant, can distinguish between opposite enantiomers, and unlike an atomic pseudoscalar measure considered in our earlier work [Phys. Rev. Lett., \textbf{133}, 268001 (2024)] does not rely on a predefined symmetry axis. For a twisted ethane model as well as several small molecules, this pseudoscalar correlates well with structural descriptors based on the continuous chirality measure. Overall, our results support response-based metrics as a physically meaningful and practically useful characterization of vibrational chirality. Importantly, while the FPC-based VCD estimate provides a useful quantifier of vibrational chirality, we show that it is a rather poor predictor of the actual molecular VCD response because the latter is strongly influenced by the (vibrational configuration-dependent) molecular electronic response.
\end{abstract}

\maketitle
\clearpage
% \quickwordcount{main} counted\\

\section{Introduction}
Chiral phonons, often considered as phonons carrying angular momentum, 
have recently attracted significant attention due to their proposed involvement in a growing number of phenomena across different fields, ranging from optics and ultrafast spectroscopy to magnetic quantum materials.\cite{Zhang2015, chen2019chiral, Zhu2018, Kim2023, Wang2024, Romao2024, Juraschek2025} 
Such motions were also recently considered as potentially instrumental in the chiral-induced spin selectivity (CISS) effect. \cite{Bloom2024, Fransson2023,Fransson2025, CISS-1, CISS-2, nuomin2026theories, Eckvahl2023} 
This recent focus on phonon chirality, primarily in the solid-state physics literature, motivates the need for a deeper understanding of the concept of chirality in molecular vibrations.

Historically, chirality in chemistry has been long formulated as a structural, time-invariant concept, e.g., 
a molecule is classified as chiral if it is not superimposable on its mirror image. 
This notion was later quantified by Avnir and co-workers \cite{avnir1995CCM, avnir2008CCM, Zabrodsky1992, c2}
by introducing the continuous chirality measure (CCM),
the minimal distance of the structure to its nearest achiral one.
Since vibrational modes can also be interpreted as generalized spatial structures, 
CCM has been applied in recent works as a vibrational chirality metric to the modes of a molecular helix \cite{Abraham2024, ethan2023quantifying} and small molecules \cite{feng2025chiral}.

An alternative route is to quantify vibrational chirality through dynamical response properties affected by the structure. 
Vibrational modes describe the response of the equilibrium geometry to a small external force, 
while chiroptical absorption/extinction encodes the response of the molecule to circularly polarized light.
The relevant response properties can be expressed as pseudoscalars,
specifically, parity-odd and time-even pseudoscalars following Barron's criterion for ``true chirality'' \cite{barron1986symmetry, barron1986true}.
As a consequence, such pseudoscalars are chirality measures that by construction are sensitive to the handness of enantiomers (same magnitude but opposite signs for opposite enantiomers), 
whereas the standard CCM is non-negative and therefore cannot distinguish the two handnesses.
Along this route, Abraham and Nitzan \cite{ethan2023quantifying} have introduced a mode chirality measure, 
an axial helicity pseudoscalar defined for each mode $k$ as the sum product of atomic linear and angular momenta vectors along a specified molecular axis $z$, that is,
\begin{align} \label{eq:helicity_axial} 
\mathscr{H}_k^{\rm Axial}  =  
\frac1{E_k} \sum_A \closedm{\vec{p}_{A}^{(k)}}_z \closedm{\vec{L}_{A}^{(k)}}_z,
% = \sum_i m_i \closedm{\vec{\nu}_i^{(k)}}_z % \closedm{R_i \times \vec{\nu}_i^{(k)}}_z.
\end{align}
(the superscript on the left denotes the axial nature of this measure and will be abbreviated as A later, in contrast to the interatomic helicity measure defined later and denoted I).
Here, $\vec{p}_{A}^{(k)}$ and $\vec{L}_{A}^{(k)}$ are the linear and angular momentum, respectively, associated with the motion of atom $A$ along the direction of the mode,
and $E_k$ is the energy of the mode. 
Note that the notation $\closedm{\px{2} \vec{\cdot}\px{3}}_z$ implies the $z$-component of a Cartesian vector, 
while $\sum_A \vec{p}_{A}^{(k)} \cdot \vec{L}_{A}^{(k)} = 0$ for all modes $k$.
The use of this axial helicity pseudoscalar as a chirality measure
was tested on twisted helical chains where a symmetry axis is clearly present \cite{ethan2023quantifying}
and on a group of small molecules \cite{feng2025chiral}  
in which the choice of axis is less obvious. 
While the axial helicity \cite{ethan2023quantifying} was found to correlate well with the corresponding CCM measure \cite{ethan2023quantifying, feng2025chiral},  the need to refer to a specific molecular axis is a clear drawback. 
Furthermore, correlating such a measure with an experimental observable will clearly enhance its credibility and usefulness. 

In this study, we extend the definition of the molecular helicity pseudoscalar to include interatomic products while dropping its axial character
and show that this measure also correlates well with the corresponding CCM measure.
Moreover, this new measure is related to a particular vibrational circular dichroism (VCD) measure of the molecule.
This measure is not the experimental VCD observable \cite{keiderling2020structure} whose quantitative magnitude is strongly affected by electron-vibration interaction,
but the solely nuclear response property calculated using the fixed partial charge (FPC) model \cite{schellman1973vibrational, deutsche1968optical, barron2009molecular}. 
This pseudoscalar is a functional of only the Born-Oppenheimer potential energy surface near equilibrium and requires no pre-defined molecular symmetry axis.
Furthermore, it can be finite even when the time-averaged angular momentum of the mode vanishes. 
We assess the performance of this inter-atom pseudoscalar by comparing against CCM chirality measures, including a CCM mode gradient measure that is introduced here for the first time. 
The correlation between this pseudoscalar measure and the vibrational CCM on the one hand, and its association with a VCD expression constructed to reflect a purely vibrational property on the other, supports its adoption as a reliable measure of vibrational chirality. 

% \clearpage
\section{VCD-inspired helicity pseudoscalar as a chirality measure \label{sec:method}}
The VCD signal is the differential absorption between left- and right- circularly polarized light in the vibrational regime of the molecular spectrum.
Its intensity can be predicted by the Rosenfeld rotational strength $R$ \cite{rosenfeld1929quantenmechanische, andrews2020physical}. 
For an isotropic sample and for the fundamental transition $1 \leftarrow 0$ of vibrational mode $k$, the Rosenfeld equation gives 
\begin{align} \label{eq:rosenfeld}
R = \textbf{Im} \closeds{\closede{0|\hat{\vec{\mu}}|1_k} 
\cdot \closede{1_k|\hat{\vec{m}}|0} },
\end{align}
resulting from interference between the electric and magnetic contributions to the transition dipole. 
Here and below an upper arrow ($\vec{\cdot}$) represents a 3-dimensional Cartesian vector. 

In the fixed partial charge (FPC) model for vibrational circular dichroism \cite{schellman1973vibrational, deutsche1968optical, barron2009molecular} proposed here as a measure of mode chirality, the dipole operators are assumed to represent the motion of atomic nuclei moving with fixed nuclear charges. 
We begin with the general expressions for the electric and magnetic dipole moments (as in Eq.~\ref{eq:rosenfeld}) of a molecule,
\begin{subequations} 
\begin{align} 
\hat{\vec{\mu}} &= \sum_i e Z_i \hat{\vec{r}}_{i}, \label{eq:dipole_moments_a} \\
\hat{\vec{m}} &= \sum_i ({e Z_i}/{2m_i}) \hat{\vec{r}}_{i} \times \hat{\vec{p}}_i.
\label{eq:dipole_moments_b}
\end{align}
\end{subequations}
The sums are over all charged particles (nuclei and electrons) in the molecule.
Here, $e$ is the electron charge and 
$eZ_i$, $m_i$, $\hat{\vec{r}}_{i}$, $\hat{\vec{p}}_i$, are the charge, mass, and position and momentum operators, respectively, of particle $i$.
Because this model is intended to describe circular dichroism in the vibrational regime of the molecular spectrum,
electron motions are disregarded and their effect is incorporated through fixed effective atomic charges, 
and the nuclear motion takes place on the adiabatic ground electronic potential surface.
This provides an approximate but physically intuitive picture of VCD
by representing the molecule as an assembly of atoms
carrying fixed charges and attributing the resulting electric and magnetic dipoles to the linear and circular motion of those charges.
These effective atomic charges, denoted below $Z_A$ for atom $A$, are obtained from population analysis based on the ground state electronic density distribution in the molecular equilibrium geometry, see for example L\"{o}wdin, \cite{szabo2012modern} Cioslowski\cite{Cioslowski1989}, Mulliken \cite{szabo2012modern} and Hirshfeld \cite{hirshfeld1977bonded}.
Because the FPC model does not track the time-evolution of the effective charges but fixes them at the values obtained at the molecular equilibrium configuration, 
the induced electromagnetic moments are determined only by the nuclear displacements and momenta,
leading to the following form of the Rosenfeld expression for the molecular optical rotational strength associated with the normal mode $k$,
\begin{align} \label{eq:R_fpc}
R_{\rm FPC}^{(1_k \leftarrow 0)} = \frac{\hbar}4 \sum_{A, B} 
e^2 Z_A Z_B \vec{\nu}_A^{(k)} \cdot \closeds{\vec{R}_{B} \times \vec{\nu}_B^{(k)}},
\end{align}
where $\vec{R}_A$ is the equilibrium position of atom $A$, 
and $\vec{\nu}_A^{(k)}$ is the displacement amplitude of atom $A$ along the $k$-th normal mode that satisfies $\sum_A m_A \vec{\nu}_A^{(k)} \cdot \vec{\nu}_A^{(k')} = \delta_{kk'}$.
Note that in Eq.~\ref{eq:R_fpc}, only terms with $A \px{-1} \neq \px{-1} B$ contribute, indicating that the electric and magnetic dipoles induced by the same atomic motion are orthogonal in the FPC model.
Furthermore, this expression can be transformed to an equivalent pairwise form 
$R_{\rm FPC}^{(1_k \leftarrow 0)} = ({\hbar}/8) \sum_{A\neq B} 
(e^2 Z_A Z_B)(\vec{R}_{B} - \vec{R}_{A}) \cdot (\vec{\nu}_B^{(k)} \times \vec{\nu}_A^{(k)}) $ (see Eq.~\ref{eq:R_pairwise} below),
which makes it explicit that Eq.~\ref{eq:R_fpc} is independent of the choice of the origin.

We thus define a chirality measure for the molecular normal mode $k$ as
\begin{align} \label{eq:helicity_inter_atom}
    \mathscr{H}_k^{\rm I} 
    & \coloneqq \sum_{A \neq B} {e^2 Z_A Z_B} \px{2}\vec{\nu}_A^{(k)} \cdot
    \closeds{\vec{R}_{B} \times  \vec{\nu}_B^{(k)}} 
    = \frac{4}{\hbar} R_{\rm FPC}^{(1_k \leftarrow 0)}
\end{align}
(the I designation denotes the interatomic nature of this helicity measure).
We note that Masiello and coworkers \cite{moser2025conservation} have pointed out the importance of intersite correlations in determining the optical chirality of a nanoplasmonic setup. 
Several more points are worth emphasizing (we note that properties (a-c) below were discussed in Ref.~\onlinecite{moser2025conservation} for the two-component momentum pseudoscalar, and are carried over to the present formulation in the molecular vibration context):
\begin{enumerate}
    \item[(a)] This pseudoscalar, Eq.~\ref{eq:helicity_inter_atom}, is completely determined by the molecular equilibrium structure and the motion in the linear regime about it.
    An alternative dimensionless pseudoscalar can be defined accordingly (see footnote).
    \footnote{A dimensionless chirality measure can be introduced corresponding to the VCD $g$-dissymmetry factor used in the chiroptical experiments, defined as the VCD intensity relative to the ordinary infrared absorption intensity. 
    Within the FPC model, this dissymmetry factor $g^{(k)} =
    (2\omega^{(k)}/c)\closedm{\sum_{A \neq B} {Z_A Z_B} \px{2}\vec{\nu}_A^{(k)} \cdot
    (\vec{R}_{B} \times  \vec{\nu}_B^{(k)})} \left/
    \closedm{\sum_{A, B} {Z_A Z_B} \px{2}\vec{\nu}_A^{(k)} \cdot \vec{\nu}_B^{(k)}} \right.$.
    However, this measure is not defined for infrared inactive vibrational modes.}
    \item[(b)] In the Appendix A, we show that $\mathscr{H}_k^{\rm I}$ is reduced to  
    $({Z^2e^2}/m^2{{E^{(k)}}})  \sum_{A \neq B} \vec{p}_{A}^{(k)} \cdot \vec{L}_{B}^{(k)}$
    for a system that consists of identical charged particles $Z_A = Z, m_A=m,\forall A$. 
    In the reduced expression, $\vec{p}_{A}^{(k)}$ and $\vec{L}_{A}^{(k)}$ are the linear and angular momentum, respectively, associated with the motion of atom $A$ along the direction of the mode with energy $E_k$ (see Appendix A).
    We note again that Eq.~\ref{eq:helicity_inter_atom} adapts the momentum pseudoscalar metric of Ref.~\onlinecite{moser2025conservation} (see Eq.~16 therein) as an explicit molecular vibrational chirality metric, and establishes its connection to the VCD rotational strength within the FPC approximation.
    \item[(c)] Compared to the axial helicity measure for mode $k$\cite{ethan2023quantifying}
    \begin{align} \label{eq:axial_mode}
    \mathscr{H}_k^{\rm A} = 
    \sum_A m_A^2 \closedm{\vec{\nu}_A^{(k)}}_z \closedm{R_A \times \vec{\nu}_A^{(k)}}_z,
    \end{align}
    the helicity measure $\mathscr{H}_k^{\rm I}$ emphasizes the inter-atom correlation between the atomic linear and angular momenta as the VCD intensity does. 
    Both mathematical forms share the same helicity motif as a product of $\vec{p}$ (parity-odd, time-odd) and $\vec{L}$ (parity-even, time-odd) vectors, so that the pseudoscalars are parity-odd and time-even.
    In addition, the per-atom weight of the contribution to the pseudoscalar is $eZ_A$ in $\mathscr{H}_k^{\rm I}$ versus $m_A$ in $\mathscr{H}_k^{\rm A}$; 
    the former weight can take either sign, while the latter is strictly positive.
    Finally, this formulation of $\mathscr{H}_k^{\rm I}$ removes the need to specify a molecular symmetry axis in $\mathscr{H}_k^{\rm A}$;
    $\sum_{A} \vec{\nu}_A^{(k)} \cdot (R_A \times \vec{\nu}_A^{(k)}) = 0$
    but $\sum_{A \neq B} \vec{\nu}_A^{(k)} \cdot (R_B \times \vec{\nu}_B^{(k)})$ is not expected to vanish in general.
    \item[(d)] A more general form of a helicity pseudoscalar based on the same core motif is 
    \begin{align} \label{eq:helicity_generalized}
    \mathscr{H}_k = \sum_{A \neq B} w_A w_B \vec{\nu}_A^{(k)} \cdot
    \closeds{\vec{R}_B \times \vec{\nu}_B^{(k)}}
    \end{align}
    where the weights $w_A, w_B$ can be functions of mass, charge, or other atomic properties. 
    An important property of this form is being 
    translationally and rotationally invariant.
    To see this, we start with Eq.~\ref{eq:helicity_generalized} and exchanging the $A \leftrightarrow B$ indices, we have
    \begin{align} \label{eq:R_pairwise}
    \mathscr{H}_k &= \sum_{A \neq B} w_A w_B \vec{\nu}_B^{(k)} \cdot
    \closeds{-\vec{\nu}_A^{(k)} \times \vec{R}_A} \nonumber \\
    & = \frac12 \sum_{A \neq B} w_A w_B \vec{\nu}_A^{(k)} \cdot
    \closedm{\closeds{\vec{R}_B - \vec{R}_A}\times \vec{\nu}_B^{(k)}}.
    \end{align}
    Because the term $\vec{R}_B - \vec{R}_A$ does not depend on the origin of the coordinate system, $\mathscr{H}_k$ is also independent of origin, i.e.,  translation-invariant.
    At the same time, for any proper rotation $U \in SO(3)$ and ${\rm det}(U) = 1$,  we have
    \begin{align}
    \mathscr{H}_k^{U} &= 
    \frac12 \sum_{A \neq B} w_A w_B \px{2}
    U \vec{\nu}_A^{(k)} \cdot
    \closedm{U \closeds{\vec{R}_B - \vec{R}_A}\times U\vec{\nu}_B^{(k)}}
    = {\rm det}(U) \mathscr{H}_k = \mathscr{H}_k,
    \end{align}
    so that the inter-atom helicity is also rotational-invariant. 
    If $U$ is instead an improper rotation (${\rm det}(U) = -1$), similar logic leads to $\mathscr{H}_k^{U} = - \mathscr{H}_k$.
    Note, however, that for the choice $w_A = m_A, \forall A$ the pseudoscalar Eq.~\ref{eq:helicity_generalized} vanishes when the normal mode $k$ satisfies the translational and rotational Eckart conditions \cite{Eckart1935,Sayvetz1939,Ferigle1953} 
    ($\sum_A m_A \vec{\nu}_A^{(k)} = 0$, $\sum_A m_A \vec{R}_A \times \vec{\nu}_A^{(k)} = 0$
    \footnote{The translational Eckart condition is origin independent.
    The rotational Eckart condition is also invariant under a shift of the coordinate origin. 
    Indeed, for an arbitrary constant vector $\vec{D}$,
    $\sum_A m_A (\vec{R}_A - \vec{D}) \times \vec{\nu}_A^{(k)} - $ 
    $\sum_A m_A \vec{R}_A \times \vec{\nu}_A^{(k)} = -\sum_A \vec{D} \times m_A \vec{\nu}_A^{(k)} = 0$ vanishes as long as the translational Eckart condition satisfies.})
    which ensures that vibrational modes do not involve translation or rotation of the rigid molecular equilibrium structure. 
    Indeed, these conditions imply that
    $\sum_{A \neq B} m_A m_B \vec{\nu}_A^{(k)} \cdot (\vec{R}_B \times \vec{\nu}_B^{(k)}) =
    \sum_{A, B} m_A \vec{\nu}_A^{(k)}
    \cdot 
     m_B (\vec{R}_B \times \vec{\nu}_B^{(k)}) =0$. 
     In quantum chemistry practice, the Eckart conditions can be enforced using the projected-Hessian procedure of Miller and co-workers \cite{miller1980reaction}.
\end{enumerate}

To end this section, we emphasize again that the chirality measure proposed above is closely related to a Rosenfeld expression evaluated for the FPC model. 
This expression is usually not a good measure of the observed molecular VCD \cite{nafie2011vibrational, Polavarapu1998}.
The latter depends on the molecular electronic structure and the associated electron-vibration interaction. 
In contrast, the FPC-based measure reflects the pure nuclear dynamics which depend on the electronic structure only through the nuclear potential surfaces and therefore is more readily identified as a property of the vibrational dynamics.
For completeness, we briefly review the electronic structure theory of VCD in Appendix B, and compare its predictions relative to those obtained from the FPC model in Section~\ref{sec:results} below.  

\section{CCM-based metrics for vibrational chirality}
As already noted, the connection (if any) between structural chirality and its dynamical manifestations is not obvious. To enhance our understanding of this connection quantitative comparisons are helpful.
The continuous chirality measure (CCM) \cite{avnir1995CCM, avnir2008CCM, Zabrodsky1992, c2} provides a quantitative metric of chirality for arbitrary geometric objects.
For a $N$-atom molecular structure descriptor, represented as $\mathbf{O}=(\vec{o}_1, \vec{o}_2, ...,\vec{o}_N)$, 
the CCM expression is defined by 
\begin{equation}  \label{eq:CCM}
    {\rm CCM}(\mathbf{O})= \frac12 \px{2} 
    {\rm min}_{\{\sigma,\mathcal{P}\}}
    \left\{1 - \frac{ \sum_{A=1}^N \vec{o}_A \cdot \sigma{\vec{o}_A}'}{\sum_{A=1}^N |\vec{o}_A|^2} \right\}, 
\end{equation}
which represents the distance between a chiral structure and its nearest non-chiral counterpart (the latter is defined as the average of an enantiomer structure and its mirror image). 
In Eq.~\ref{eq:CCM}, $\mathbf{O}$ is a $3N$-dimensional object defined by a group of sites $A$, 
$\vec{o}_A$ are the Cartesian site position vectors, the operator $\sigma$ applies a mirror plane reflection and the operator $\mathcal{P}$ permutes over atom indices $A$ of the same element type. 
$\{{\vec{o}_A}'\}$ are the components of the structure obtained by $\mathcal{P}$ applied to $\{\vec{o}_A\}$.
The minimization is taken over all possible mirror planes and all such permutations.

The definition (\ref{eq:CCM}) leaves us with the freedom to define the structure descriptor $\mathbf{O}$ for which the CCM is evaluated.
Below we examine three choices, denoted CCM$_{1/2/3}$.  
\begin{enumerate}
\item[(1)] The standard molecular CCM \cite{Alvarez2005,Carreras2019}, here denoted CCM$_1$, uses $\mathbf{O} = (\vec{R}_1, \vec{R}_2, \cdots, \vec{R}_N)$, the atom positions of the equilibrium molecular geometry to provide a chirality
measure for the molecular equilibrium configuration. 
\item[(2)] A similar measure for molecular normal modes was introduced in our recent works \cite{feng2025chiral, ethan2023quantifying}.
For the CCM of mode $k$, here denoted CCM$_2^{(k)}$,
we use $\mathbf{O} = (\vec{\nu}_1^{(k)}, \cdots, \vec{\nu}_N^{(k)} )$,
the Cartesian atomic displacements associated with mode $k$.
CCM$_2$ = $N_{\rm modes}^{-1} \sum_k {\rm CCM}_2^{(k)}$ is the average of this measure over all modes. 
Following Refs.~\onlinecite{feng2025chiral, ethan2023quantifying} we also consider partial averages over modes in certain frequency ranges, as detailed in Sec.~\ref{sec:results} below.
\item[(3)]
Here, we also examine another possible choice for the CCM of normal modes, the CCM mode gradient, defined as
\begin{align} \label{eq:ccm_mode_grad}
{\rm CCM}_3^{(k)} = \lim_{\eta \to 0} \left|
\frac{ 
{\rm CCM} (\mathbf{R} + \eta \px{2} \delta\mathbf{R}^{(k)}) - 
{\rm CCM}(\mathbf{R} - \eta \px{2}  \delta \mathbf{R}^{(k)})}{2\eta} \right|
\end{align}
where $\mathbf{R} = (\vec{R}_1, \vec{R}_2, \cdots, \vec{R}_N)$ is the vector
describing the molecular equilibrium configuration, and
$\delta\mathbf{R}^{(k)}
= (\vec{\nu}_1^{\,(k)}, \cdots, \vec{\nu}_N^{\,(k)} )$ is
the vector of the mode displacements.
More precisely, Eq.~\ref{eq:ccm_mode_grad} is defined as a symmetrized gradient, i.e., the average of the left and right derivatives if they differ. 
\footnote{This can happen, for example, in the case of an achiral molecule (thus ${\rm CCM}(\textbf{R})=0$) and CCM$(\mathbf{R} + \eta \px{2} \delta\mathbf{R}^{(k)})$ is linear with respect to $|\delta \textbf{R}^{(k)}|$ for a particular mode $k$; the left and right CCM mode derivatives have opposite signs, whereas the symmetrized gradient is 0. 
The symmetrized gradient is chosen to avoid mathematical ambiguity when the gradient is not well-defined.}
Note that for CCM$_3^{(k)}$ only the absolute magnitude is physically meaningful since the overall sign of $\nu^{(k)}$ or $\delta\mathbf{R}^{(k)}$ is arbitrary. 
Unlike the dynamical measure defined earlier, these three CCM metrics cannot distinguish left- and right-enantiomers.
\end{enumerate}
In studying the dynamical implications of molecular chirality, an interesting and potentially important question concerns the correlation between the chirality of the molecular equilibrium structure and that of its nuclear motions. 
In this regard, we note that the 
molecules with achiral equilibrium structures (i.e., CCM$\textbf{(R)} = 0$) cannot have chiral modes characterized as such by the CCM$_3^{(k)}$ measure.
This observation is consistent with the observation in Ref.~\onlinecite{feng2025chiral} 
that CCM depends quadratically on the infinitesimal twist about the equilibrium structure.
However, these modes may be identified as chiral according to other measures, such as CCM$_2^{(k)}$ or $\mathscr{H}^{\rm I}_k$, as we shall see in Sec.~\ref{sec:results}.

\newpage
\begin{figure}[!htb]
    \centering
    \includegraphics[width=0.5\textwidth]{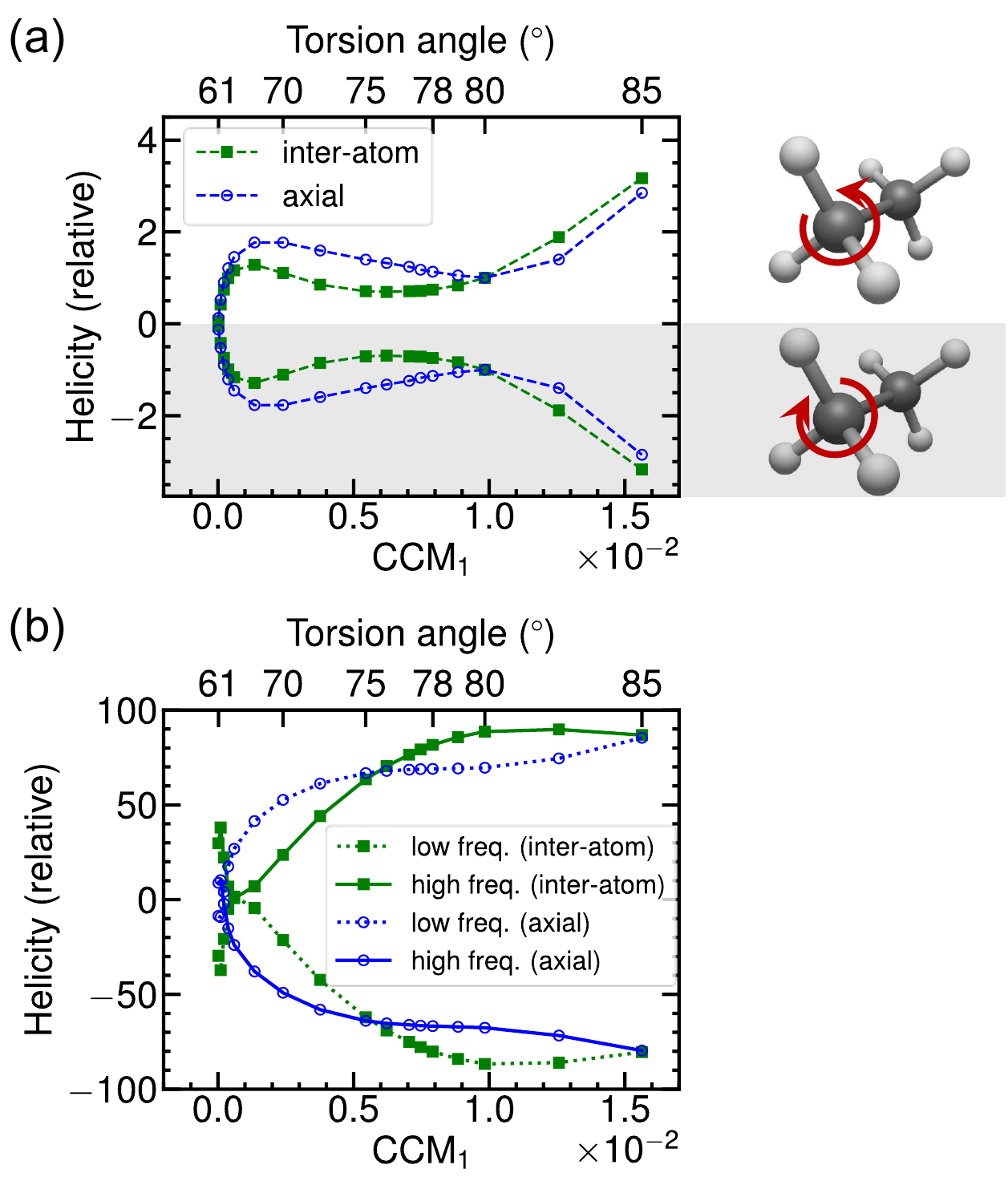}
    \caption{Inter-atom (green) and axial (blue) helicity pseudoscalars as chirality measures for a twisted ethane molecule. 
    The twist generates a series of geometries with increasing \ch{H-C-C-H} dihedral angle from 60$^\circ$ to 85$^\circ$.
    \textbf{(a)} For each twisted geometry, these helicity pseudoscalars (averaged over all vibrational modes) are plotted against the structural continuous chirality measure (CCM$_1$). 
    The results in the white- and gray-shaded area correspond to the distortion of the ethane configuration towards two opposite twist directions shown in schematic illustrations.
    \textbf{(b)} The same helicity pseudoscalars are calculated separately for the groups of high frequency and low frequency modes (see text), plotted against CCM$_1$
    for the right-handed twisted ethane molecule (as in the white-shaded area of panel (a)).
    To enable a dimensionless comparison, the helicity pseudoscalars in both panels are reported relative to their values at the twisted geometry with a dihedral angle of 80$^\circ$
    which are $\bar{\mathscr{H}}^{\rm A} =$
    6.4 $\times 10^{-5}$ amu $\cdot \; a_0$ $= 5.6 \times 10^{-42}$ kg$\cdot$m for the axial helicity and $\bar{\mathscr{H}}^{\rm I} =$
    5.5 $\times 10^{-6}$ $e^2 a_0$ $\cdot$ amu$^{-1} = 4.5 \times 10^{-27}$ C$^2$ m kg$^{-1}$ for the inter-atom helicity with $a_0$ the Bohr radius.
    % $1$ amu $\cdot a_0$ $= 8.8 \times 10^{-38}$ kg$\cdot$m for the axial helicity, 
    % and $1$ $e^2 a_0$ $\cdot$ amu$^{-1} = 8.2 \times 10^{-22}$ C$^2$ m kg$^{-1}$ for the inter-atom helicity with $a_0$ the Bohr radius.
    The vibrational modes of ethane are partitioned into high- (solid) and low- frequency groups (dotted), and each group consists of half of the modes.
    \label{fig:ethane}}
\end{figure}

\section{Results and Discussions \label{sec:results}}
Fig.~\ref{fig:ethane} compares the two helicity pseudoscalars for the model of a twisted ethane molecule studied in Ref.~\onlinecite{feng2025chiral}. 
In this model, the positions of the C atoms are held fixed while one \ch{-CH3} group is forced to rotate about the \ch{C-C} bond. 
Starting from the centrosymmetric, staggered conformation, the \ch{-CH3} group is incrementally twisted in both directions (indicated by the white- and gray-shaded regions in Fig.~\ref{fig:ethane}(a)). 
For each molecular geometry along this torsional coordinate, 
the instantaneous vibrational modes are obtained from the molecular Hessian at that structure, corresponding to a local harmonic approximation to the torsional potential energy surface.
The atomic partial charges are assigned with L\"{o}wdin population analysis \cite{szabo2012modern}.
Computational details of the needed electronic structure calculations are reported in Appendix C.
The inter-atom and axial\cite{feng2025chiral} pseudoscalars are then evaluated with Eqs.~\ref{eq:helicity_inter_atom} and~\ref{eq:helicity_axial} in Sec.~\ref{sec:method}.
The helicities displayed in Fig.~\ref{fig:ethane}(a) are the average, over all molecular normal modes ($\bar{\mathscr{H}} = N_{\rm modes}^{-1} \sum_{k} \mathscr{H}_{k}$), of these pseudoscalars plotted as a function of geometrical CCM, i.e., CCM$_1$, of the instantaneous geometries (see also Appendix C for computational details).
As expected, the structural chirality increases, as quantified by the CCM$_1$ values, with increasing dihedral angle between the two \ch{C-H} bonds.
The white- and gray-shaded areas in this panel show results obtained by twisting the molecular structure (changing the dihedral angle) in opposite directions, with the atomic positions and the mode displacement vectors reflected across the midpoint plane perpendicular to the C-C bond.
In this example, the partial charge on the C atom only changes by $<0.01e$ over the entire twist process, % ($\delta- \in [-0.45, -0.44]$) 
therefore, the inter-atom helicity results reflect mostly the variation in the instantaneous mode vectors.
Similar trends are shown by both helicity pseudoscalars even though the weights used for the atomic contribution to the two measures 
($\propto m_A$, and $m_{\rm C} / m_{\rm H} = 12$ in the axial pseudoscalar, 
and $\propto Z_A$, and $Z_{\rm C} / Z_{\rm H} = -3$ in the inter-atom pseudoscalar) 
are quite different.
Also note that 
the values of pseudoscalars reverse sign when changing the twist direction (going from white-shaded to gray-shaded regions of the plot) while the CCM$_1$ results do not. The results mark a clear distinction; the helicity pseudoscalars distinguish left- and right-enantiomers while the CCM metric cannot.

Fig.~\ref{fig:ethane}(b) focuses on the frequency dependence of this behavior. 
As in Ref.~\onlinecite{feng2025chiral}, the vibrational modes are divided into two frequency groups of equal size, $\{G_{\rm l}, G_{\rm h}\}$, according to their vibrational frequencies. 
Then the averaged helicities of the low- and high-frequency groups, 
$\bar{\mathscr{H}}_{\rm l/h} = (N_{\rm modes}/2)^{-1} \sum_{k\in G_{\rm l}/G_{\rm h}} \mathscr{H}_{k}$ are reported in solid and dotted lines, respectively, against the CCM$_1$ values.
We see that the high- and low-frequency group averages of the inter-atom helicity exhibit opposite trends (therefore opposite signs) as the torsion angle increases,
\footnote{In a forthcoming work we address the origin and implications of this observation by examining the relation between the chemical nature of the vibrational modes and their helicity\cite{claudia}}
consistent with the behavior of the axial helicity pseudoscalar reported previously \cite{ethan2023quantifying, feng2025chiral}.
Note that the molecular distortion expressed in the chirality measure of the subgroups of low- and high-frequency modes (Fig.~\ref{fig:ethane}b) is far greater than that expressed in the chirality measures averaged over all modes (Fig.~\ref{fig:ethane}a), as seen by comparing the corresponding vertical scales in these figures.
The comparatively small magnitude of the all-mode averages reflects substantial cancellation of the helicities between the low- and high-frequency groups, whose helicities contribute with opposite signs.
It is also interesting to note that the two helicity pseudoscalars $\bar{\mathscr{H}}^{\rm I}$ and $\bar{\mathscr{H}}^{\rm A}$ carry opposite signs for the frequency-group averages $\bar{\mathscr{H}}_{\rm l/h}$ (but the same sign for overall average helicity $\bar{\mathscr{H}}$).

\clearpage
\begin{figure}[!htb]
    \centering
    \includegraphics[width=0.75\textwidth]{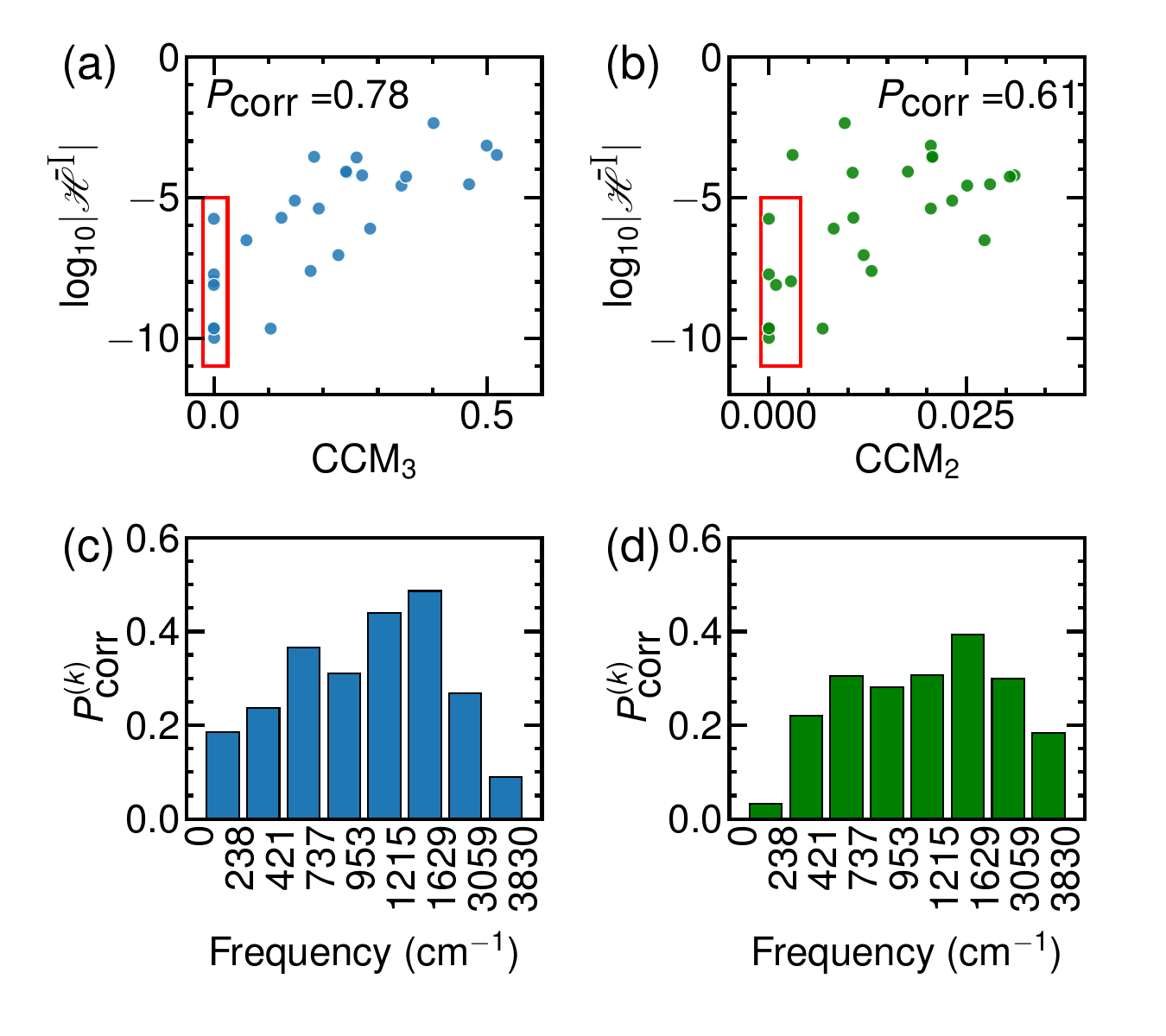}
    \caption{\textbf{(a-b)}  Correlation between the logarithms of the dynamical-response-based (interatomic helicity pseudoscalar $\bar{\mathscr{H}}^{\rm I}$) and the structure-based vibrational chirality measures (CCM$_3$ in panel (a), blue and CCM$_2$ in panel (b), green).
    The scattered dots correspond to the mode-averaged values for individual molecules and the resulting Pearson correlation coefficient ($P_{\rm corr}$) are also shown. 
    Red boxes denote molecules with achiral equilibrium geometries.
    Panels \textbf{(c-d)} show these correlations within individual groups of the vibrational modes, grouped according to their frequencies. 
    Each group consists of the same number of modes.
    The superscript $(k)$ of $P_{\rm corr}^{(k)}$ indicates that the correlation is evaluated for the individual mode $k$, rather than averaged over all modes of a given molecule in (a-b).
    The horizontal ticks indicate the edges of the frequency groups. 
    \label{fig:molecule_set}}
\end{figure}

Fig.~\ref{fig:molecule_set} examines the correlation between the inter-atom helicity measure 
and two CCM metrics when used as a quantifier of mode-resolved vibrational chirality.
The test set contains 26 small molecules
that were used in Ref.~\onlinecite{feng2025chiral} to assess the correlations between vibrational chirality measures.
Note that not all molecules in the set have clearly defined symmetry axis, and the axial helicity used in Ref.~\onlinecite{feng2025chiral} was not always applicable for such molecules.
The inter-atom helicity definition, however, is applicable in all cases.
The vibrational normal modes and the atomic partial charges (from L\"{o}wdin population analysis) were obtained with the quantum chemistry calculations of the molecular equilibrium geometries (see Appendix C for the details of the electronic structure calculations).
In particular, the projected-Hessian procedure of Miller {\it et al.} \cite{miller1980reaction} is employed to ensure that the vibrational modes have zero mixing with the global translations and rotations of the rigid molecule. 
Three mode-resolved chirality measures, including
the inter-atom helicity measure ($\mathscr{H}^{\rm I}_k) $, as well as the two structural, mode-resolved CCM measures (CCM$_{2}^{(k)}$, CCM$_{3}^{(k)}$) are evaluated with Eq.~\ref{eq:helicity_inter_atom} and Eq.~\ref{eq:CCM}
(see Appendix C for the details of CCM calculations).
The vibrational chirality measures averaged over all modes for each molecule, namely, $\bar{\mathscr{H}}^{\rm I} = N_{\rm modes}^{-1} \sum_{k} \mathscr{H}_{k}^{\rm I}$ and 
CCM$_j = N_{\rm modes}^{-1} \sum_{k} {\rm CCM}_j^{(k)}$ are compared in the scatter plots in Fig.~\ref{fig:molecule_set}(a) (blue, for $\bar{\mathscr{H}}^{\rm I}$ v.s. CCM$_3$) and (b) (green, for $\bar{\mathscr{H}}^{\rm I}$ v.s. CCM$_2$), respectively.
The Pearson correlation coefficients, $P_{\rm corr}$,
\footnote{For two data series $\{X_j\}$ and $\{Y_j\}$,
$P_{\rm corr} = {\sum_{j} (X_j-\bar X)(Y_j-\bar Y)}/
({\sqrt{\sum_{j} (X_j-\bar X)^2}\sqrt{\sum_j (Y_j-\bar Y)^2}}$.}
between the logarithm of $\bar{\mathscr{H}}^{\rm I}$ and the structural chirality measures CCM$_2$ and CCM$_3$ are reported to quantify the linear correlation between dynamical and structural chirality measures.
Fig.~\ref{fig:molecule_set}(c-d) investigates the frequency dependence of the correlation between the dynamical and structural chirality measures for the vibrational modes in the small molecules dataset.
A total of 645 vibrational modes of the same set of molecules as in Fig.~\ref{fig:molecule_set}(a-b) are partitioned evenly into eight groups according to their frequencies (i.e., same number of modes in each group up to divisibility constraints). 
We then evaluate the Pearson correlation between the logarithm of inter-atom helicity magnitude, $\log_{10}|\mathscr{H}_k^{\rm I}|$, and the mode-resolved structural measures,  CCM$_2^{(k)}$ and CCM$_3^{(k)}$, respectively, for modes within each frequency group and denoted as $P_{\rm corr}^{(k)}$.
Here in (c-d), $P_{\rm corr}^{(k)}$ is evaluated mode by mode, irrespective of the molecule from which each mode originates. 
The horizontal axis reports the frequency percentiles, and the vertical axis gives the corresponding linear correlation coefficient computed within each frequency group for $\log_{10}|\mathscr{H}_k^{\rm I}|$ versus CCM$_3^{(k)}$ (blue, panel (c)) and 
$\log_{10}|\mathscr{H}_k^{\rm I}|$ versus  CCM$_2^{(k)}$ (green, panel (d)) respectively.

The following observations can be made:
\begin{enumerate}
\item[\emph{(i)}] 
When averaged over all modes of a given molecule, 
the dynamical and the structural chirality measures exhibit mutual correlations.
The observation is quantitatively supported by the correlation coefficients ($P_{\rm corr}=0.78$, panel (a)) between the dynamical/$\log_{10}|\bar{\mathscr{H}}^{\rm I}|$ and structural/CCM$_3$ measures 
(a value of 0 corresponds to no linear dependency and a value of $\pm1$ corresponds to perfect linear dependency).
The correlation between the helicity pseudoscalar with CCM$_2$ (panel (b)) is slightly weaker ($P_{\rm corr}=0.61$).
Finally, although not shown here, we note that the linear correlation between the mode-averaged absolute of the helicity pseudoscalar, $\log_{10} (N_{\rm modes}^{-1} \sum_k |{\mathscr{H}}_k^{\rm I}|)$ and CCM$_3$ is markedly weaker ($P_{\rm corr}=0.53$). 
This observation reinforces the point that the relative signs of the mode helicities carry physically meaningful information.
\item[\emph{(ii)}] The results displayed in panels (a-b) show that 
for molecules with the achiral equilibrium structures (dots in the solid red box, both (a) and (b) panels), 
CCM$_3^{(k)}$ is strictly 0 for all mode $k$ thus has a mode-average of 0.
Non-zero, albeit small values are observed
with CCM$_2^{(k)}$ or $\mathscr{H}^{\rm I}_k$ measures.
We conclude that there is no unique way to assign definite chirality to a vibrational mode as the assignment depends on the measure used. 
Similarly, molecules with chiral equilibrium structures can possess vibrational modes with different degrees of chirality, depending again on the mode-resolved measure used.
\item[\emph{(iii)}] The correlation between the considered dynamical and structural chirality measures displayed in panels (c-d)
appears to be mode-dependent and shows a non-monotonic frequency dependence.
This mode-specific correlation, 
evaluated for $\log_{10}|\mathscr{H}_k^{\rm I}|$ v.s. both CCM$_3^{(k)}$ and CCM$_2^{(k)}$ for each mode $k$
peaks at an intermediate frequency range.
In addition, $P_{\rm corr}^{(k)}$ is somewhat weaker than $P_{\rm corr}$ presented in panels (a-b),
where the correlation is evaluated molecule by molecule.
This frequency dependence is likely associated with motion type, as lower-frequency modes are often collective motions such as bending and torsion modes, while higher-frequency modes are often stretching modes.
\end{enumerate}

\newpage
\renewcommand{\arraystretch}{1.5}
\begin{table}[!h]
\centering
\begin{tabular}{l|cccc} 
\hline
{$|P_{\rm corr}^{(k)}|^2$} & \ch{H2O2} & \ch{CH2N2} & \ch{N2H4} & \ch{(C6H5)_2} \\ 
\hline
$\mathscr{H}_k^{\rm I}$	v.s. $R_k$	&	0.37	&	0.89	&	0.47	&	0.57	\\
\hline
CCM$_2^{(k)}$ v.s. $|R_k|$  &	0.04	&	0.10	&	0.01	&	0.04	\\
\hline
CCM$_3^{(k)}$ v.s. $|R_k|$  & 0.80 & 0.11 & 0.51 & 0.02	\\
\hline
\end{tabular}
\caption{ 
The linear correlation coefficients ($|P_{\rm corr}^{(k)}|^2$) between the Rosenfeld rotational strength ($R$) from the electronic structure calculations (see Appendix B)
versus the inter-atom helicity measure (Eq.~\ref{eq:helicity_inter_atom}),
and the vibrational CCM measures.
The correlation is calculated for each molecule on a mode-to-mode basis for the four molecules presented in Fig.~\ref{fig:molecules}.
For example, in the first data row, we report $|P_{\rm corr}^{(k)}|^2 = 0.37$ for the linear correlation between the inter-atom helicity pseudoscalar measure $\mathscr{H}_k^{\rm I}$ versus the electronic structure expression of the rotational strength $R_k$, evaluated over the six vibrational modes of \ch{H2O2}. \label{table:values}
}
\end{table}
\begin{figure}[!htb]
    \centering
    \includegraphics[width=0.5\linewidth]{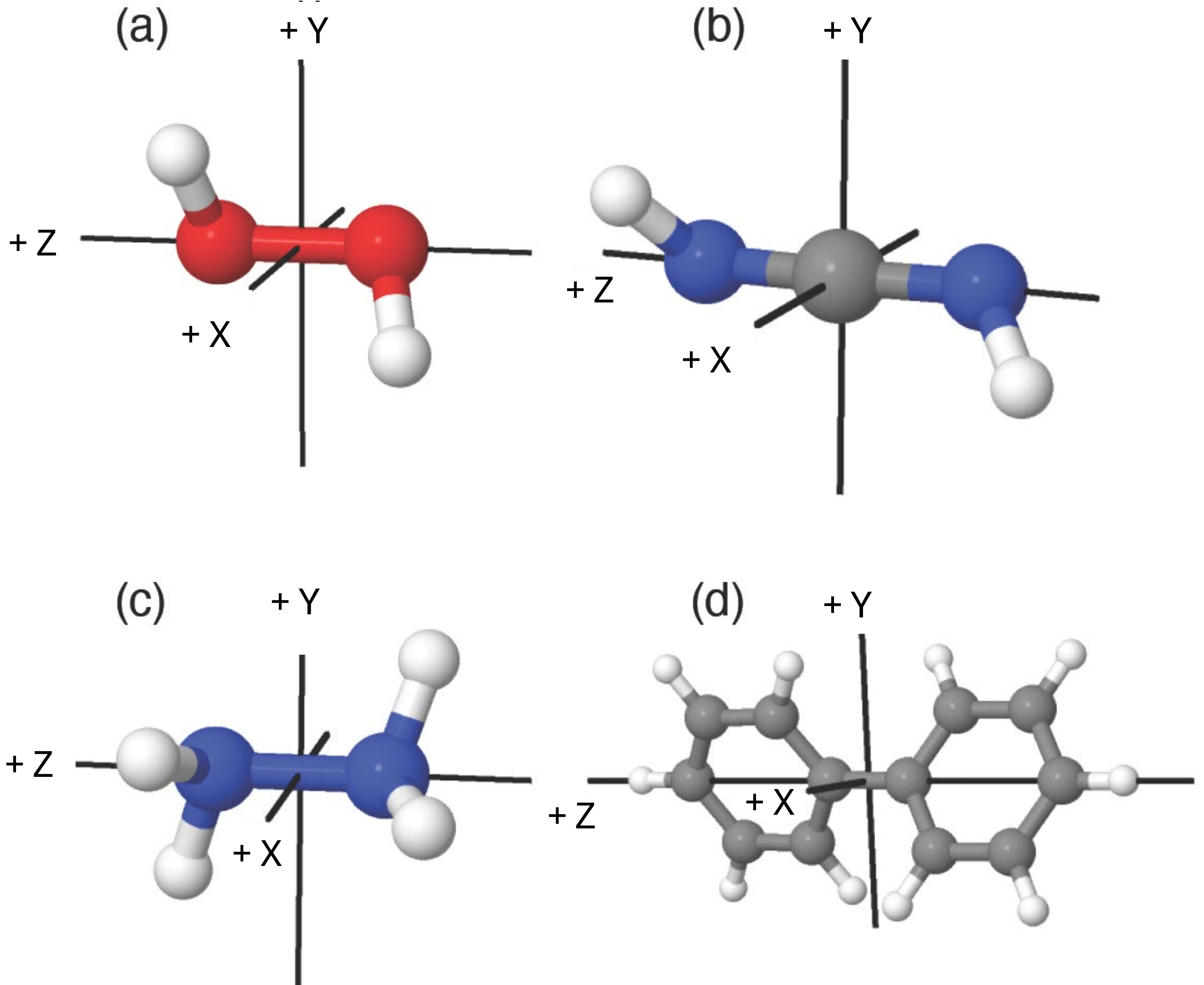}
    \caption{Molecular structures of (a) hydrogen peroxide (\ch{H2O2}), (b) methanediimine (\ch{CH2N2}), (c) hydrazine (\ch{N2H4}), and (d) biphenyl (\ch{(C6H5)_2}). In all cases, the molecular axis is aligned with the $z$ axis.}
    \label{fig:molecules}
\end{figure}

These results in Figs.~\ref{fig:ethane} and \ref{fig:molecule_set} suggest that the dynamical and structural chirality measures are aligned, but the degree of alignment varies across different types of vibrations. 
The inter-atom helicity pseudoscalar, a response-based metric, appears to be a useful metric for characterization of vibrational chirality. It should be emphasized, however, that vibrational chirality defined as such by mode CCM or by this helicity pseudoscalar is not by itself a predictor of vibrational optical dichroism. 
To see this, we consider the set of molecules displayed in Fig.~\ref{fig:molecules}, and calculate for each of their normal modes
their $\mathscr{H}_k^{\rm I}$ and CCM$^{(k)}$ values, as well as the Rosenfeld rotational strengths ($R_k$) obtained with electronic structure calculations (see Appendix C for computational details).
These data sets are then used to calculate Pearson correlation strengths (see footnote 44 on page 14) and the results are provided in Table \ref{table:values}. Modest correlations are seen between $\mathscr{H}_k^{\rm I}$  (essentially the rotational strength calculated in the FPC approximation, $\propto R_{\rm FPC}$ in Eq.~\ref{eq:R_fpc}) and $R_k$,
while the correlation of the latter with the vibrational CCM metrics is substantially poorer.
Moreover, since the presence of a linear correlation does not exclude a difference in magnitude, we also compare $\mathscr{H}_k^{\rm I}$ with $R_k$ directly in Table S1 of the Supplementary Material and found that the FPC approximation can underestimate the rotational strength with deviations of order of magnitude.
This underestimation may be understood by 
considering the limitation of the FPC model as a predictor of the rotational strength,
which may be illustrated with the VCD selection rules inferred from point group analysis.
For the rotational strength to be non-zero, from Eq.~\ref{eq:R_ES} (in Appendix B), we see that the following conditions must be satisfied simultaneously. 
\begin{equation}
     \frac{\partial }{ \partial Q_k} \langle \Psi_g | \hat{\vec{\mu}} | \Psi_g\rangle \neq 0 \quad\text{and}\quad
     \frac{\partial }{ \partial \dot{Q}_k} \langle \Psi_g | \hat{\vec{m}} | \Psi_g\rangle \neq 0
\end{equation}
For these to hold, the direct product of the irreducible representations should contain the totally symmetric representation,\cite{Cotton1991,Polavarapu1998} i.e., 
$\Gamma_Q \otimes \Gamma_{\Psi_g} \otimes \Gamma_{O_\alpha} \otimes \Gamma_{\Psi_g} \in A$, where $O=\hat{\vec{\mu}}, \hat{\vec{m}}$,
$\alpha$ specifies the Cartesian index, 
and we have used $\Gamma_{\dot{Q}}=\Gamma_Q$.
Since the ground state wavefunction of a closed shell system belongs to the totally symmetric irreducible representation (A), the above conditions imply that 
\begin{equation} \label{eq:group_theory}
    \Gamma_Q \otimes \Gamma_{\hat \mu_\alpha} \in A \quad \text{and} \quad
    \Gamma_Q \otimes \Gamma_{\hat m_\alpha} \in A.
\end{equation}
Eq.~\ref{eq:group_theory} implies that not only the irreducible representation of the normal mode matters.
In other words, what determines whether a specific molecular vibration originates a VCD signal (in an isotropic sample) depends not only on the characteristics of the nuclear vibrations, 
but also on those of the electric and magnetic dipole moments involving 
both electronic and nuclear contributions.
The neglect of electronic response 
beyond a localized charge approximation in the simple FPC model (and in other measures of vibrational chirality) leads to the underestimation of the VCD intensity seen above. 
\footnote{Approaches like charge flow and bond moment models, \cite{abbate1981charge, polavarapu1983comparison, barron2009molecular} which represent improvements over the FPC approximation, provide more quantitatively accurate VCD predictions but will not necessarily improve the estimate of nuclear modes chirality.}
Regardless of this limitation, the rotational strength evaluated from the FPC model provides a useful metric for the chirality of molecular vibrations. It should be emphasized that neither of these measures is directly associated with another commonly addressed and potentially important phenomenon---phonons carrying finite average angular momentum. 

\section{Summary}
This work has aimed to examine possible metrics for the chirality of vibrational normal modes. 
To this end, we developed an inter-atom helicity pseudoscalar as a quantitative measure of mode chirality, motivated by an approximate expression for the vibrational circular dichroism (VCD) intensity within the fixed partial charge (FPC) model. 
This pseudoscalar is translationally and rotationally invariant, distinguishes opposite enantiomers, and is generally applicable to molecular systems lacking a clear symmetry axis. 
Furthermore, it correlates well with the structure-based continuous chirality measure (CCM) of the vibrational modes as well as with the CCM that characterizes the chirality of the underlying molecular equilibrium structure. 
It is important to note that this helicity pseudoscalar is a measure of correlation between linear and angular momentum, and does not necessarily indicate a phonon that carries non-zero average angular momentum. 
The distinction between chiral phonons and phonons that carry angular momentum was recently made in Ref.~\onlinecite{Juraschek2025}.

We also conclude that there is no unique way to quantify vibrational chirality. Accordingly, we examined several measures and illustrated their mutual correlations with molecular examples in this study. 
These measures quantify the degree to which improper rotation symmetry breaking is expressed in different molecular properties; 
CCM$_1$ characterizes this symmetry breaking in equilibrium geometries, 
CCM$_2$ characterizes its expression in the vibrational mode displacement patterns,
and CCM$_3$ characterizes how this symmetry breaking varies along one vibrational mode displacement near the equilibrium geometry. 
Finally, the developed inter-atom helicity pseudoscalar measures an enantiomer-sensitive correlation between linear and angular atomic motions in the displacements of individual vibrational modes.
These measures are built based on distinct properties, and their mutual correlations are by themselves significant. Overall, we suggest the application of vibrational chirality measures suited to the chemistry contexts of interest.

While providing a useful metric for vibrational chirality, this helicity pseudoscalar is not, by itself, a reliable predictor of VCD intensity because the latter does not reflect pure vibrational chirality, 
but is also affected by the dependence of electronic electric and magnetic dipoles on vibrational coordinates, and its full dependence on the molecular electronic structure cannot be accounted for by models that assume fixed nuclear charges.
While the inter-atom helicity pseudoscalar may serve as a useful tool in future studies of chiral vibrations, including chiral phonons, 
it remains to be seen whether the inter-atom helicity pseudoscalar can serve as a mode-resolved descriptor of the enantiomer-odd vibrational contribution in chiral transport.

\section*{Acknowledgement}
This research is supported by the
Air Force Office of Scientific Research under Award No.
FA9550-23-1-0368.
C.C. acknowledges support from the Spanish Ministry of Science, Innovation and Universities–Agencia Estatal de Investigaci\'on through Grant No. PID2024-159213NA-I00 and from the Beatriz Galindo Programme (BG23/00024).
The authors acknowledge Dr. David J. Masiello and Dr. Wenxiang Ying for helpful discussions.

\appendix
% \setcounter{figure}{0}
% \setcounter{table}{0}
% \makeatletter
% \renewcommand{\thefigure}{A\arabic{figure}}
% \renewcommand{\thetable}{A\arabic{table}}

\section{Quantities expressed in the normal mode coordinates}
In this Appendix, we provide details on the expressions in the normal-mode coordinates that are used in the main text. 
Much of the content follows what has been discussed previously, including the textbook \cite{barron2009molecular} and research articles \cite{ethan2023quantifying, feng2025chiral}.

We follow the convention to use the term normal mode (i.e., a 3-dimensional vector $\vec{\nu}_A^{(k)}$ for the $k$-th harmonic mode, $A$-th atom) throughout the article for the vibrational eigenmodes that are orthonormal to each other as $\sum_A m_A \vec{\nu}_A^{(k)} \cdot \vec{\nu}_A^{(k')} = \delta_{kk'}$ for non-degenerate modes.
We also make the time dependence of the vibrational motion explicit by writing the time-dependent mode amplitude as $X^{(k)}(t) = X^{(k)} \cos (\omega^{(k)} t) $, $X^{(k)} \in \mathbb{R}$ with $\omega^{(k)}$ the mode frequency,
and $\closede{E^{(k)}}_t = \closedm{\omega^{(k)}}^2 \closedm{X^{(k)}}^2 / 2$.
It follows that the mode displacements and their associated momenta are
\begin{alignat}{2}
\delta \vec{r}_{A}^{(k)} &= {\vec{r}_{A}^{(k)}  - \vec{R}_{A}} \px{2} &=& \px{2} X^{(k)} \cos (\omega^{(k)} t) \px{2}  \vec{\nu}_{A}^{(k)}, \nonumber \\ 
\vec{p}_{A}^{(k)} &= {m_A \delta \dot{\vec{r}}_A}^{(k)} &=& \px{2} - \omega^{(k)} X^{(k)} \sin (\omega^{(k)} t)  m_A \vec{\nu}_A^{(k)},
\end{alignat}
and the angular momenta are
\begin{align}
%{\rm conj} 
\vec{L}_{B}^{(k)} = \px{2} \vec{r}_{B}^{(k)} \times \vec{p}_{B}^{(k)}
= - \omega^{(k)} X^{(k)} \sin (\omega^{(k)} t) \vec{R}_{B} \times  m_B \vec{\nu}_B^{(k)}.
\end{align}
Note that the time average $\closede{\vec{L}_{B}^{(k)}}_t = 0$ for such sinusoidal oscillatory motions, 
indicating that the vibrational modes carry zero net angular momentum.
The conclusion is general regardless of the center of the origin to evaluate the vector $\vec{R}_B$.
However, helicity terms such as $\vec{p}_{A}^{(k)}  \cdot \vec{L}_{B}^{(k)}$ may exhibit a non-zero time average.

Since $\vec{p}_{A}^{(k)}  \cdot \vec{L}_{A}^{(k)} 
\propto \vec{\nu}_A^{(k)} \cdot (\vec{R}_{A} \times \vec{\nu}_A^{(k)}) 
= \vec{R}_{A} \cdot (\vec{\nu}_A^{(k)} \times \vec{\nu}_A^{(k)}) = 0$, the previous works chose to use the $z$-component of 
the inner product (Eq.~\ref{eq:helicity_axial}) to characterize the chirality, i.e.,
\begin{align} 
\mathscr{H}_k^{\rm Axial}  &=  
\frac1{\closede{E^{(k)}}_t} \sum_A \closede{\closedm{\vec{p}_{A}^{(k)}}_z  \closedm{\vec{L}_{A}^{(k)}}_z}_t 
= \sum_A m_A^2 \closedm{\vec{\nu}_A^{(k)}}_z 
\closedm{\vec{R}_A \times \vec{\nu}_A^{(k)}}_z
% =\frac{1}{E_k}\sum_i p_{k,i}^z (p_{k,i}^y x_i-p_{k,i}^x y_i),
\end{align}
Following the same procedure, we also have 
\begin{align} \label{eq:m-helicity}
\frac1{\closede{E^{(k)}}_t} \closede{\sum_{A,B} \vec{p}_{A}^{(k)} \cdot \vec{L}_{B}^{(k)}}_t
&= \sum_{A,B} m_A m_B \px{2} \vec{\nu}_A^{(k)} \cdot
\closeds{\vec{R}_B \times \vec{\nu}_B^{(k)}} \nonumber \\
&= \sum_{A \neq B} m_A m_B \px{2} \vec{\nu}_A^{(k)} \cdot
\closeds{\vec{R}_B \times \vec{\nu}_B^{(k)}}.
\end{align}
Therefore, for a system consisting of identical charged particles $Z_A = Z, m_A=m,\forall A$, 
the inter-atom helicity pseudoscalar
$\mathscr{H}_k^{\rm I} = (Z^2e^2 /m^2 \closede{E^{(k)}}_t) \sum_{A \neq B} \closede{\vec{p}_{A}^{(k)} \cdot \vec{L}_{B}^{(k)}}_t $.
Specifically, in the system of identical particles, Eq. A4 vanishes for vibrational modes that satisfy the Eckart conditions (also see point (d) in the main text Sec.~\ref{sec:method}).
For simplicity, we drop the time average notation $\closede{\cdot}_t$ in the main text whenever no ambiguity arises.

\section{VCD from electronic structure calculations}
The interatomic mode chirality measure introduced in Section~\ref{sec:method} is based on the form assumed by the Rosenfeld expression for the molecular rotational strength in the FPC model. 
It is often the case, however, that matrix elements in Eq.~\ref{eq:rosenfeld} are dominated by nuclear coordinate dependence of the electronic dipole moments, the evaluation of which requires a full electronic structure calculation.
For completeness, 
we briefly review the electronic structure theory of VCD and
consider the extent to which the FPC model can approximate its VCD prediction for the fundamental (1 $\leftarrow$ 0) transition of a molecular vibrational mode.

Actual molecular VCD intensities are predicted using magnetic field perturbation theory, \cite{stephens1985theory} nuclear velocity perturbation, \cite{nafie2011vibrational, nafie2000circular, Nafie1992, buckingham1987velocity} or the recently developed phase-space approach \cite{duston2024phase, bian2026phase}.
Most electronic structure calculations start from the rotational strength for the fundamental transition of mode $k$ expressed as
\begin{equation}
    R^{(1_k \leftarrow 0)} = \textbf{Im} \closedm{\langle \Psi_g 0 | \hat{\vec{\mu}} | \Psi_g 1_k \rangle 
    \cdot \langle \Psi_g 1_k| \hat{\vec{m}} |\Psi_g 0 \rangle}
\end{equation}
where $\Psi_g$ denotes the electronic ground state wavefunction.
By expanding the electric and magnetic dipole moment operators, $\hat{\vec{\mu}}$ and $\hat{\vec{m}}$, in terms of the harmonic normal coordinates $Q_k$ about the ground state equilibrium geometry (denoted by {\rm eq}), the expression can be recast as \cite{nafie1983vibronic, stephens1985theory, nafie2011vibrational, buckingham1987velocity}
\begin{equation} \label{eq:R_ES}
    R^{(1_k \leftarrow 0)} = \hbar \px{2} \textbf{Im}
    \Bigg(\frac{\partial }{ \partial Q_k} \langle \Psi_g | \hat{\vec{\mu}} | \Psi_g\rangle\Bigg)_{\rm eq} \cdot
    \Bigg(\frac{i}{2}  \frac{\partial }{ \partial \dot{Q}_k} \langle \Psi_g |  \hat{\vec{m}} | \Psi_g\rangle \Bigg)_{\rm eq}.
\end{equation}
Furthermore, Eq.~\ref{eq:R_ES} can be expressed in a compact form as \cite{Bak1993} 
\begin{equation}
    R^{(1_k \leftarrow 0)} = \hbar \px{2} \textbf{Im}
    [\tilde{\boldsymbol{P}}_k \cdot \tilde{\boldsymbol{M}}_k]
\end{equation}
where the vectors 
$\tilde{\boldsymbol{P}}_k$ and $\tilde{\boldsymbol{M}}_k$ are defined in terms of their $\beta$ Cartesian components
\begin{align} 
    \tilde{P}_{\beta,k} \equiv \sum_{A,\alpha} P_{\alpha \beta}^A \closedm{\vec{\nu}_A^{(k)}}_\alpha, \px{8}
    \tilde{M}_{\beta,k} \equiv \sum_{A,\alpha} M_{\alpha \beta}^A \closedm{\vec{\nu}_A^{(k)}}_\alpha
\end{align}
with $\closedm{\vec{\nu}_A^{(k)}}_\alpha$ being the $\alpha$ Cartesian components of the normal mode vectors $\vec{\nu}_A^{(k)}$ introduced in Sec.~\ref{sec:method}.
$P_{\alpha \beta}^A$ and $M_{\alpha \beta}^A$ are the atomic polar tensor (APT) and the atomic axial tensor (AAT), respectively.

The atomic polar tensor (APT) is given by
\begin{equation} \label{eq:APT_AAT}
    P_{\alpha \beta}^A \equiv  E_{\alpha \beta}^A + N_{\alpha \beta}^A  
\end{equation}
while the atomic axial tensor (AAT) is given by
\begin{equation}
    M_{\alpha \beta}^A \equiv I_{\alpha \beta}^A + J_{\alpha \beta}^A  
\end{equation}
In both expressions, the two terms correspond to the electronic and nuclear contributions, respectively, and are defined by \cite{Bak1993}
\begin{equation}
\begin{split}
    E_{\alpha \beta}^A \equiv \Big( \frac{\partial }{\partial R_{A\alpha}} \langle \Psi_g | \hat{\mu}_\beta^E | \Psi_g \rangle \Big)_{\rm eq}, \px{8}
    N_{\alpha \beta}^A \equiv Z_A e\delta_{\alpha \beta}
\end{split}
\end{equation}
and 
\begin{equation}
\begin{split}
    I_{\alpha \beta}^A \equiv \frac{i}{2}\Big( \frac{\partial }{\partial \dot{R}_{A\alpha}} \langle \Psi_g | \hat{m}_\beta^E | \Psi_g \rangle \Big)_{\rm eq}, \px{8}
    J_{\alpha \beta}^A \equiv \frac{i}{4}Z_A e\sum_\gamma \varepsilon_{\alpha \beta \gamma}R_{A\gamma}^{\rm eq}
\end{split}
\end{equation}
where $\alpha$, $\beta$, $\gamma$ denote Cartesian components, $\varepsilon_{\alpha \beta \gamma}$ is the Levi-Cevita tensor, and $R_{A \gamma}^{\rm eq}$ is the equilibrium position of atom $A$.
The superscript $E$ of the dipole operators refers to the electronic contribution only.

Within the Born-Oppenheimer approximation, $\langle \Psi_g | \hat{m}_\beta^E | \Psi_g\rangle = 0$ for a non-degenerate ground state (hence real wavefunctions) because the electronic magnetic dipole operator is Hermitian and purely imaginary. 
Thus, the above expression for $I_{\alpha \beta}^A$ must be evaluated beyond the Born-Oppenheimer approximation.  
Application of magnetic field perturbation framework (Stephens formalism) \cite{Amos1987, stephens1985theory} leads to
\begin{equation}
    I_{\alpha \beta}^A = \hbar \Bigg \langle 
    \Bigg( \frac{\partial \Psi_g}{\partial R_{A\alpha}} \Bigg) 
    \Bigg|
    \Bigg( \frac{\partial \Psi_g}{\partial H_\beta} \Bigg)
    \Bigg \rangle_{{\rm eq}, \boldsymbol{H}=0}
\end{equation}
where derivatives of the electronic wavefunction with respect to an external magnetic field $H_\beta$ are required. Electronic structure VCD calculations within the Stephens formalism thus require the use of London or gauge independent atomic orbitals (GIAOs), whereby the atomic orbitals are multiplied by a complex phase that explicitly depends on an external magnetic field. \cite{Bak1995} Importantly, using London atomic orbitals in these types of calculations ensures gauge origin independent results.\cite{Bak1993} 

\section{Computational Details}
\begin{figure}[!htb]
    \centering
    \includegraphics[width=0.5\textwidth]{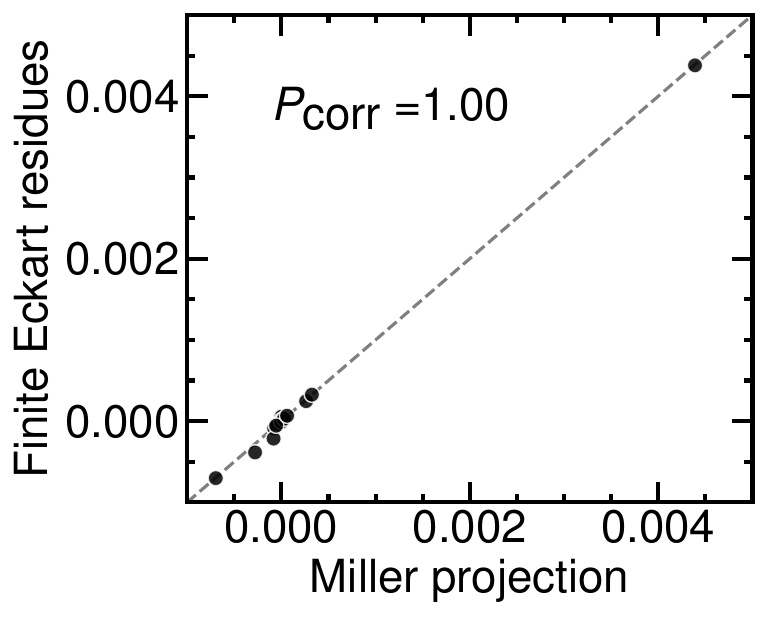}
    \caption{The helicity pseudoscalar $\mathscr{H}_k^{\rm I}$, Eq.~\ref{eq:helicity_inter_atom}, evaluated using the vibrational modes from standard GAMESS quantum chemistry output (i.e., modes with small but finite residues to the Eckart conditions, vertical axis) v.s. the same helicity pseudoscalar evaluated using the vibrational modes with precisely zero translational and rotational mixing (horizontal axis) for the 26 molecules studied in Fig.~2. A Pearson correlation of $1.00$ shows that this projected-Hessian step does not influence the helicity results.
    \label{fig:miller_projection}
    }
\end{figure}

Electronic structure calculations for the results in \autoref{fig:ethane} and \autoref{fig:molecule_set} are performed with GAMESS software, version 2019 (R1) \cite{GAMESS}. 
The Hessian calculations to obtain the vibrational frequencies and the mode vectors, as well as the population analysis to assign the atomic partial charges, are performed at DFT B3LYP/6-31G* level \cite{
becke1993density, lee1988development}. 
In Fig.~\ref{fig:ethane}, the geometries of the twisted ethane molecule are obtained by gradually increasing the \ch{H-C-C-H} dihedral angle from the equilibrium geometry.
For the results in Fig.~\ref{fig:molecule_set}, 
the geometry optimization for the small molecules dataset uses a tolerance of 10$^{-5}$ Hartree/Bohr
for the root mean square gradient.
For SCF iterations, convergence is reached when the density change between two cycles is less than 10$^{-6}$.
The projected-Hessian procedure \cite{miller1980reaction} was also performed with the GAMESS software;
as shown in Fig.~\ref{fig:miller_projection}, it does not alter the helicity results.
Molecular structures are visualized using Avogadro software. \cite{Hanwell2012}

The electronic structure calculations for the results in \autoref{table:values} are performed at the Hartree-Fock level using the 6-31G* basis set with the Dalton2020.1 quantum chemistry program. \cite{Aidas2014,daltonprogram} Vibrational circular dichroism rotational strengths are computed using the Dalton implementation, which employs London atomic orbitals and follows the magnetic field perturbation formalism developed by Stephens. \cite{Bak1993,Bak1994} The effective atomic charges required to calculate the FPC rotational strengths, $R_{\rm FPC}$, were estimated from the atomic polar tensor of each atom, according to $Z_A =  \sum_\alpha P_{\alpha\alpha}^A/3$ (see also Eq.~\ref{eq:APT_AAT}). \cite{Cioslowski1989} 

We employ the Cosymlib library \cite{Alemany2021, cosymlibdoc} to perform CCM evaluations (Eq.~\ref{eq:CCM}) of the equilibrium and perturbed molecular geometries, and of the vibrational normal modes obtained from the GAMESS or Dalton calculations.
In Eq.~\ref{eq:CCM}, the CCM is defined by choosing the mirror symmetry group $C_s$ as the reference point-group symmetry in the continuous symmetry measure expression, without imposing higher improper rotational symmetries $S_n (n>1)$. We validate this assumption in Eq.~\ref{eq:CCM} with the test molecule set shown in Fig.~\ref{fig:molecules}.
In addition, we note that Ref.~\onlinecite{feng2025chiral} has concluded that accounting for atomic connectivity
when applying permutations does not affect the CCM$_1$ values for the test set of 26 small molecules considered in this work.
For the CCM mode gradient CCM$_3$ results reported in Fig.~\ref{fig:molecule_set},
we use a finite difference gradient with $\eta = 0.1, 0.05, 0.025$ but observed no significant dependence of these results on this choice.
   
% \input{main.bbl}
% \bibliography{ref.bib}
% \nocite{*}

% References
%aipnum4-2.bst 2019-01-14 (MD) hand-edited version of apsrev4-1.bst

\end{document}

\title{Supplementary Material for ``Chiral vibrational modes and vibrational circular dichroism''}
\author{Xuecheng Tao$^\dagger$}
\email{xuechengtao@gmail.com}
\affiliation{Department of Chemistry, University of Pennsylvania, Philadelphia, Pennsylvania 19104, United States}
\author{Cl\`audia Climent$^\dagger$}
\email{claudiacliment@ub.edu}
\affiliation{Departament de Ci\`encia de Materials i Qu\'imica F\'isica and 
Institut de Qu\'imica Te\`orica i Computacional (IQTCUB), Universitat de Barcelona, 
Martí i Franqu\`es 1, 08028 Barcelona, Spain}
\author{Ethan Abraham}
\affiliation{Department of Chemistry, Massachusetts Institute of Technology, Cambridge, Massachusetts 02139, USA}
\author{Jichen Feng}
\affiliation{Department of Chemistry, University of Pennsylvania, Philadelphia, Pennsylvania 19104, United States}
\author{Abraham Nitzan}
\email{anitzan@sas.upenn.edu}
\affiliation{Department of Chemistry, University of Pennsylvania, Philadelphia, Pennsylvania 19104, United States}
\affiliation{Department of Physical Chemistry, Tel Aviv University, Tel Aviv 6997801, Israel}

\freefootnote{$^\dagger$ Equal contribution.}

% \date{\today}
\maketitle
\newpage

\setcounter{figure}{0}
\setcounter{table}{0}
\makeatletter
\renewcommand{\thefigure}{S\arabic{figure}}
\renewcommand{\thetable}{S\arabic{table}}

\section*{The FPC approximation to the rotational strength}

\begin{table}[!h]
    \centering                                  
    \caption{\label{tab:example} The rotational strength from the electronic structure calculation and the FPC approximation for the molecules in Fig.~3. 
    The values of $R$ and $R_{\rm FPC}$ are compared up to a global sign common to all vibrational modes.
    In these dimer-like molecules, symmetric and antisymmetric normal modes can have similar frequencies but opposite helicities. This is illustrated, for instance, by modes 1 and 2, the symmetric and asymmetric \ch{O-H} stretching, of the \ch{H2O2} molecule. \\
    Alt text: Table listing vibrational modes for the molecules in Fig.~3. The FPC approximation for the rotational strength ($R_{\rm FPC}$) is compared with values from electronic structure calculations for each mode. } 
    \begin{tblr}{
      colspec = {
        Q[l,wd=0.12\textwidth]
        Q[l,wd=0.07\textwidth]
        Q[r,wd=0.13\textwidth]
        Q[r,wd=0.20\textwidth]
        Q[r,wd=0.20\textwidth]
        Q[r,wd=0.08\textwidth]
      },
      row{1} = {font=\bfseries},
      rowsep = 1pt,
    }
        \hline
        molecule & mode & frequency (cm$^{-1}$) & $R$ ($10^{-44} \; {\rm esu}^2 \; {\rm cm}^2$) 
        & $R_{\rm FPC} \propto \mathscr{H}^{\rm I}_k$ ($10^{-44} \; {\rm esu}^2 \; {\rm cm}^2$)   
        &$P_{\rm corr}^{(k)}$ \\ 
        \hline
        \SetCell[r=6]{halign=l, valign=h}{\ch{H2O2}}
        % \multirow{6}{*}{\ch{H2O2}}
        &    1    &    4077.54    &    -18.572    &    69.133    & 
        % \multirow[b]{6}{*}{0.60}  
        \SetCell[r=6]{halign=r, valign=f}{0.60}   \\
        &    2    &    4076.48    &    6.970    &    -69.569    &    \\
        &    3    &    1632.64    &    -22.224    &    -23.601    &    \\
        &    4    &    1496.19    &    25.396    &    18.899    &    \\
        &    5    &    1145.03    &    -4.610    &    0.079    &    \\
        &    6    &    407.83    &    232.670    &    93.092    &    \\
        \hline                                   

        \SetCell[r=9]{halign=l, valign=h}{\ch{CH2N2}}
        % \multirow{9}{*}{\ch{CH2N2}}
        &    1    &    3822.69    &    -52.819    &    -154.522    &   
        % \multirow[b]{9}{*}{0.94}
        \SetCell[r=9]{halign=r, valign=f}{0.94} \\
        &    2    &    3818.27    &    50.240    &    157.528    &    \\
        &    3    &    2358.72    &    10.416    &    -0.860    &    \\
        &    4    &    1390.39    &    0.813    &    1.091    &    \\
        &    5    &    1028.67    &    -122.851    &    -262.159    &    \\
        &    6    &    1020.96    &    128.607    &    265.477    &    \\
        &    7    &    836.69    &    296.976    &    354.246    &    \\
        &    8    &    576.32    &    -46.947    &    -87.440    &    \\
        &    9    &    574.60    &    12.121    &    61.344    &    \\
        \hline                                   

        \SetCell[r=12]{halign=l, valign=h}{\ch{N2H4}}
        % \multirow{12}{*}{\ch{N2H4}}
        &    1    &    3820.93    &    -13.691    &    20.047    &  
        \SetCell[r=12]{halign=r, valign=f}{0.69} \\
        % \multirow[b]{12}{*}{0.69}   \\
        &    2    &    3814.21    &    -3.493    &    -21.133    &    \\
        &    3    &    3712.94    &    0.114    &    5.311    &    \\
        &    4    &    3701.60    &    18.369    &    -3.990    &    \\
        &    5    &    1873.17    &    -12.776    &    -4.393    &    \\
        &    6    &    1856.57    &    9.981    &    3.118    &    \\
        &    7    &    1470.70    &    67.744    &    0.405    &    \\
        &    8    &    1436.88    &    -73.770    &    2.147    &    \\
        &    9    &    1226.61    &    15.658    &    4.301    &    \\
        &    10    &    1115.01    &    -164.809    &    -28.778    &    \\
        &    11    &    980.23    &    176.295    &    23.057    &    \\
        &    12    &    476.48    &    -64.100    &    -7.549    &    \\
        \hline                                  

        \SetCell[r=15]{halign=l, valign=h}{\ch{(C6H5)2}}
        % \multirow{12}{*}{\ch{(C6H5)2}}
        &    1    &    3388.67    &    0.000    &    0.000    &    \\
        &    2    &    3387.14    &    0.036    &    -0.008    &    \\
        &    3    &    3380.43    &    76.409    &    4.808    &    \\
        &    4    &    3377.57    &    -116.732    &    -4.214    &    \\
        &    5    &    3373.49    &    -0.001    &    0.000    &    \\
        &    6    &    3369.94    &    -0.765    &    0.081    &    \\
        &    7    &    3361.57    &    10.044    &    0.728    &    \\
        &    8    &    3358.82    &    0.087    &    -1.397    &    \\
        &    9    &    3352.66    &    0.000    &    0.000    &    \\
        &    10    &    3351.84    &    0.342    &    0.007    &    \\
        \hline
    \end{tblr}
\end{table}                                    

\setcounter{table}{0}

\begin{table}[!htbp]                                    
    \centering                                   
    \caption{(Continued.) The rotational strength from the electronic structure calculation and the FPC approximation for the molecules in Fig.~3. 
    The values of $R$ and $R_{\rm FPC}$ are compared up to a global sign common to all vibrational modes. }                                    
    \begin{tblr}{
      colspec = {
        Q[l,wd=0.12\textwidth]
        Q[l,wd=0.07\textwidth]
        Q[r,wd=0.13\textwidth]
        Q[r,wd=0.20\textwidth]
        Q[r,wd=0.20\textwidth]
        Q[r,wd=0.08\textwidth]
      },
      row{1} = {font=\bfseries},
      rowsep = 1pt,
    }
        \hline
        molecule & mode & frequency (cm$^{-1}$) & $R$ ($10^{-44} \; {\rm esu}^2 \; {\rm cm}^2$) & $R_{\rm FPC} \propto \mathscr{H}^{\rm I}_k $ ($10^{-44} \; {\rm esu}^2 \; {\rm cm}^2$)
        & $P_{\rm corr}^{(k)}$ \\ 
        \hline
        \SetCell[r=50]{halign=l, valign=h}{\ch{(C6H5)2}}
        &    11    &    1817.60    &    0.000    &    0.000    &     
        \SetCell[r=50]{halign=r, valign=f}{0.76} \\ % \multirow[b]{40}{*}{0.76} \\
        &    12    &    1807.00    &    -1.166    &    0.021    &    \\
        &    13    &    1788.18    &    3.519    &    -1.843    &    \\
        &    14    &    1770.81    &    -3.341    &    2.018    &    \\
        &    15    &    1684.98    &    0.000    &    0.000    &    \\     
        &    16    &    1658.98    &    -7.641    &    0.047    &  \\
        &    17    &    1625.91    &    -29.750    &    -0.508    &    \\
        &    18    &    1595.86    &    32.337    &    0.250    &    \\
        &    19    &    1481.29    &    2.065    &    0.560    &    \\
        &    20    &    1465.84    &    -3.578    &    -0.451    &    \\
        &    21    &    1407.25    &    0.000    &    0.000    &   \\
        &    22    &    1339.13    &    1.584    &    0.252    &    \\
        &    23    &    1326.73    &    1.148    &    0.021    &    \\
        &    24    &    1307.22    &    0.000    &    0.000    &    \\
        &    25    &    1302.53    &    -0.506    &    0.009    &    \\
        &    26    &    1225.65    &    2.913    &    -0.008    &    \\
        &    27    &    1208.81    &    0.460    &    0.055    &    \\
        &    28    &    1186.64    &    1.972    &    0.708    &    \\
        &    29    &    1184.36    &    -1.788    &    -0.160    &    \\
        &    30    &    1152.79    &    -0.863    &    -0.014    &    \\
        &    31    &    1133.41    &    0.000    &    0.000    &    \\
        &    32    &    1124.32    &    -0.274    &    -1.372    &    \\
        &    33    &    1124.26    &    1.473    &    1.063    &    \\
        &    34    &    1104.43    &    0.000    &    0.000    &    \\
        &    35    &    1102.38    &    1.008    &    1.266    &    \\
        &    36    &    1100.21    &    -1.280    &    -1.259    &    \\
        &    37    &    1089.36    &    0.000    &    0.000    &    \\
        &    38    &    1084.86    &    -0.086    &    -0.003    &    \\
        &    39    &    1052.44    &    -0.288    &    -0.282    &    \\
        &    40    &    1034.06    &    11.619    &    -0.210    &    \\
        &    41    &    960.05    &    0.000    &    0.000    &    \\
        &    42    &    958.55    &    -4.400    &    -0.186    &    \\
        &    43    &    878.35    &    -74.298    &    -5.210    &    \\
        &    44    &    835.43    &    133.634    &    5.958    &    \\
        &    45    &    807.56    &    0.000    &    0.000    &    \\
        &    46    &    781.11    &    -84.822    &    -0.460    &    \\
        &    47    &    779.72    &    86.592    &    0.055    &    \\
        &    48    &    685.75    &    -0.330    &    -0.012    &    \\
        &    49    &    673.98    &    -0.114    &    -0.009    &    \\
        &    50    &    668.57    &    3.039    &    0.037    &    \\
        &    51    &    609.18    &    -21.955    &    -0.565    &    \\
        &    52    &    553.20    &    58.618    &    0.038    &    \\
        &    53    &    461.44    &    0.000    &    0.000    &    \\
        &    54    &    454.36    &    0.163    &    0.001    &    \\
        &    55    &    391.19    &    -9.399    &    0.286    &    \\
        &    56    &    330.51    &    0.000    &    0.000    &    \\
        &    57    &    298.48    &    0.799    &    -0.147    &    \\
        &    58    &    130.92    &    10.427    &    -0.089    &    \\
        &    59    &    104.71    &    -25.048    &    0.134    &    \\
        &    60    &    67.09    &    0.000    &    0.000    &    \\
        \hline
    \end{tblr}                                 
\end{table}                                    

\section*{Mode-by-mode comparison of VCD rotational strength and CCM measures}
\textcolor{white}{ }

\begin{figure}[!h]
    \includegraphics[width=0.8\linewidth]{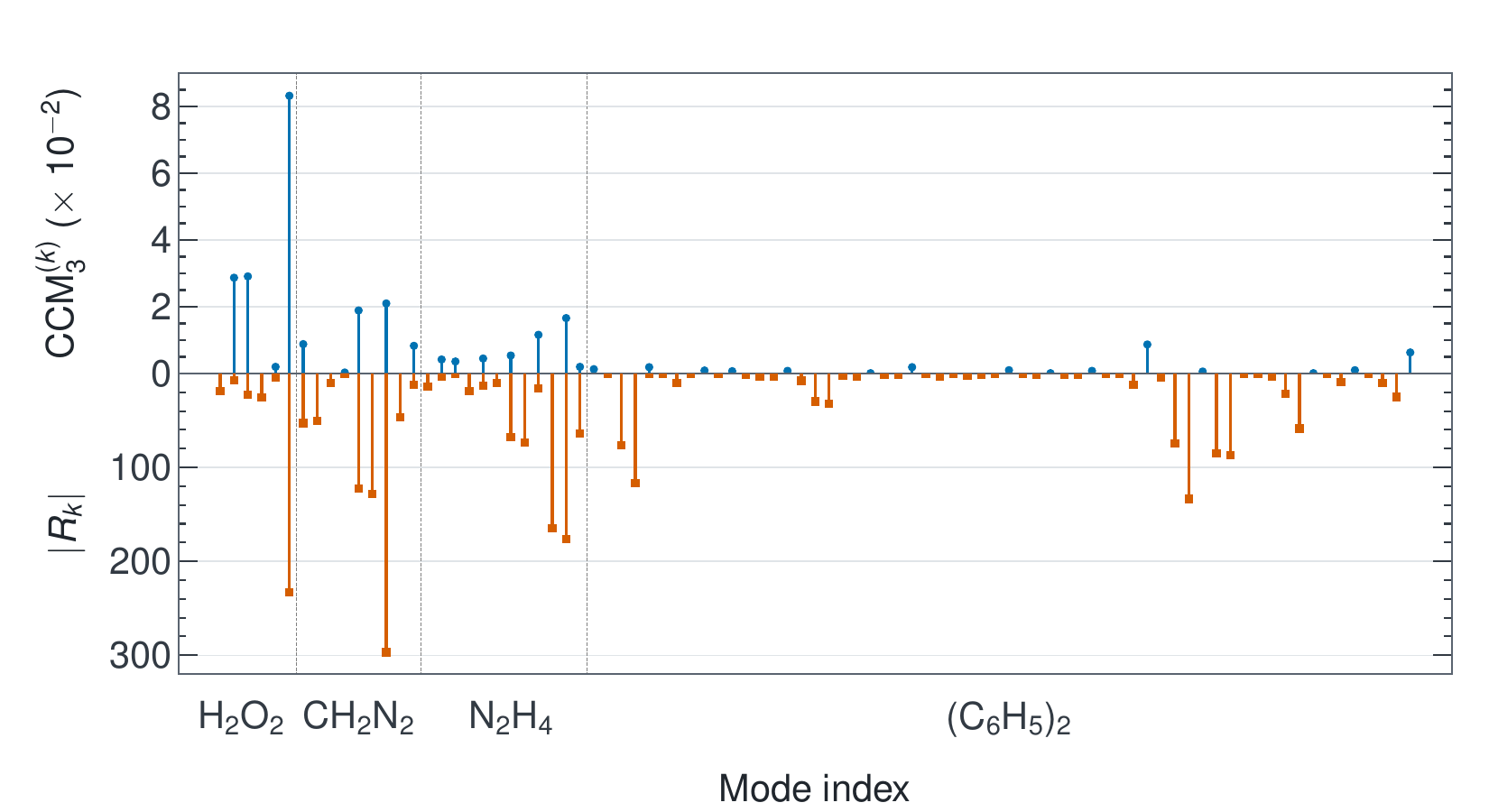}
    \caption{Mode-by-mode comparison between the $\textrm{CCM}_3^{(k)}$ measure (always non-negative) and the magnitude of VCD rotational strengths $R_k$ for each vibrational mode, $k$, for the molecules shown in Fig.~3. The vertical dotted lines separate the vibrational modes belonging to different molecules. \\
    Alt Text: Mirrored bar plot comparing the non-negative $\textrm{CCM}_3^{(k)}$ structural chirality measure (top, scaled) against the magnitude of the VCD rotational strength $|R_k|$ (bottom) for the vibrational modes of the molecules in Fig.~3. }
\end{figure}
\begin{figure}[!h]
    \centering
    \includegraphics[width=0.8\linewidth]{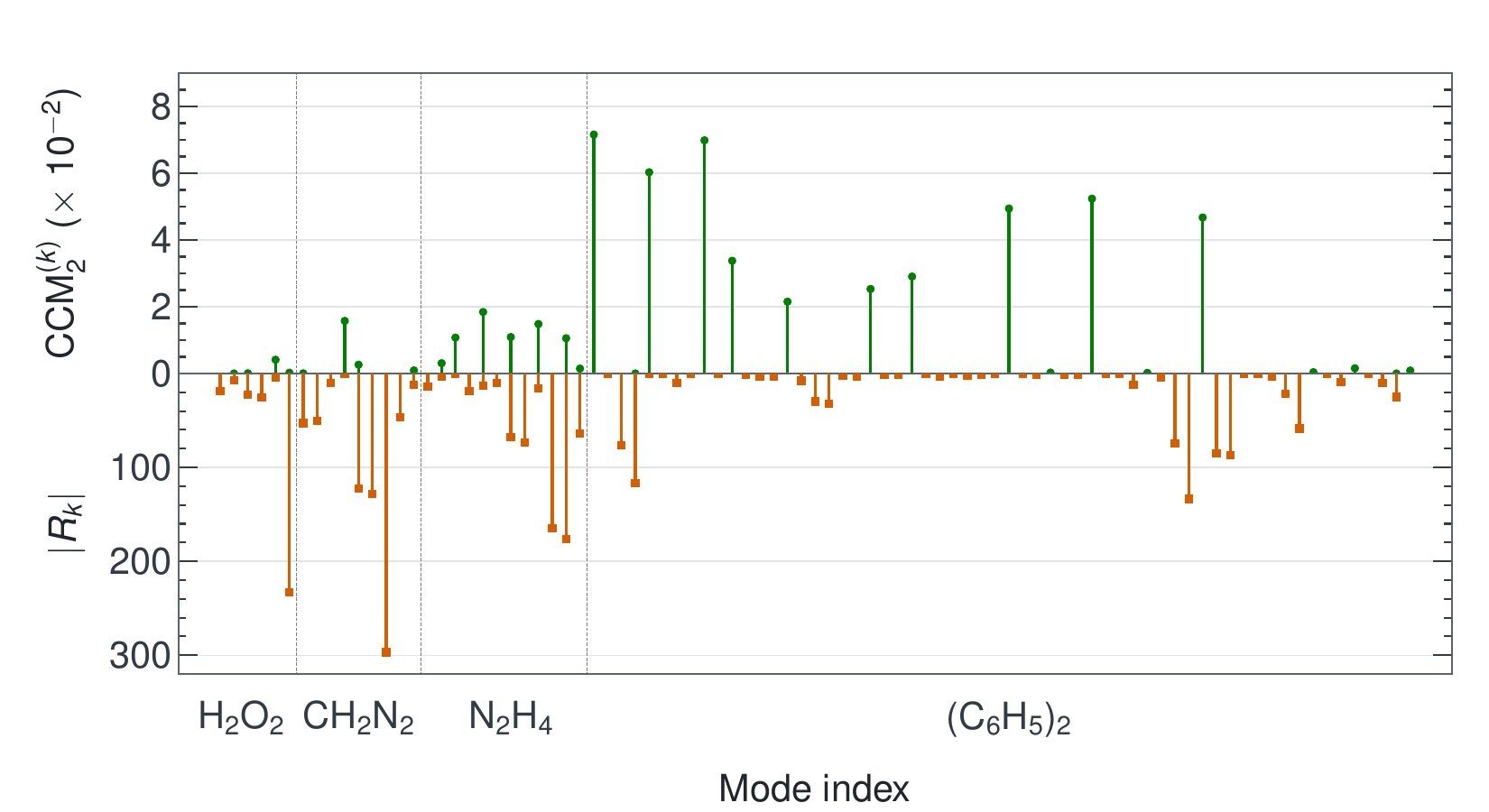}
    \caption{Mode-by-mode comparison between the $\textrm{CCM}_2^{(k)}$ measure (always non-negative) and the magnitude of VCD rotational strengths $R_k$ for each vibrational mode, $k$, for the molecules shown in Fig.~3. The vertical dotted lines separate the vibrational modes belonging to different molecules. \\
    Alt Text: Mirrored bar plot comparing the non-negative $\textrm{CCM}_2^{(k)}$ structural chirality measure (top, scaled) against the magnitude of the VCD rotational strength $|R_k|$ (bottom) for the vibrational modes of the molecules in Fig.~3. }
\end{figure}